# Modelling the inelastic constitutive behaviour of multi-layer spiral strands. Comparison of hysteresis operator approach to multi-scale model.

Davide Manfredo[1,2], Mohammad Ali Saadat[3], Vanessa Dörlich[1], Joachim Linn[1], Damien Durville[3], Martin Arnold[2]

[1] Mathematics for the Digital Factory
Fraunhofer Institute for Industrial Mathematics (ITWM)
Fraunhofer-Platz 1, 67663, Kaiserslautern, Germany
[davide.manfredo,vanessa.doerlich,joachim.linn]@itwm.fraunhofer.de

[2] Institute of Mathematics,
Martin Luther University Halle-Wittenberg,
Theodor-Lieser-Str. 5, 06120 Halle (Saale), Germany
martin.arnold@mathematik.uni-halle.de

[3] Laboratoire de Mécanique Paris-Saclay
Université Paris-Saclay, Centrale Supélec
ENS Paris-Saclay, CNRS, LMPS
3 rue Joliot-Curie, Gif-sur-Yvette, 91190, France
[mohammad-ali.saadat,damien.durville]@centralesupelec.fr

**ABSTRACT**

The simulation of inelastic effects in flexible slender technical devices has become of increasing interest in the past years. Different approaches have been considered depending on the effects relevant for the specific application. Recently, a mixed stress-strain driven computational homogenisation has been proposed to model the dissipative nonlinear bending response of spiral strands subjected to axial force. In this study, we propose two different approaches, namely a rheological model and a data-based greybox model, to predict the cyclic response of these strands using only their monotonic response. In the first approach, a system of so-called bending springs and sliders is used to model different contributions to the bending stiffness of the strands. The data-based approach makes use of mathematical tools called hysteresis operators. The Prandtl-Ishlinskii operator plays a relevant role in modelling the input-output relation in phenomena showing hysteretic behaviour and can be expressed as a weighted superposition of elementary stop operators. Comparing the two approaches leads to a better understanding and an explicit physical interpretation of the parameters of a specific class of hysteresis operator models.

## 1 INTRODUCTION

Spiral strands are commonly used in various engineering applications and consist of a central wire surrounded by multiple helically wound wires, forming a helical pattern. Due to the helical geometry, the spiral strands exhibit axial-torsional coupling, and also dissipative nonlinear bending behaviour when subjected to tensile force. To characterise the bending behaviour of these strands, the mixed stress-strain driven computational homogenisation [1] can be used.

Computational homogenisation aims to numerically determine the effective material properties of a homogeneous medium equivalent to a heterogeneous medium. The effective material properties are obtained by solving a boundary value problem (BVP) on a microsample known as a representative volume element (RVE), where all heterogeneities are explicitly taken into account. It should be emphasised that a major advantage of computational homogenisation is that it does not require any a priori assumption of the effective material properties. In the case of spiral strands, the heterogeneous medium consists of wires arranged in multiple layers, the homogeneous medium is represented as a single beam, and the RVE is a short sample of the strand. As explained in [1], due to the dependence of the bending behaviour of the spiral strands on the axial stress and the geometric coupling of bending curvature and axial stress, a mixed stress-strain driven homogenisation is used for spiral strands in contrast to the conventional strain-driven homogenisation at finite strain. It should be mentioned that 1D kinematically enriched beam element is used to model individual wires in the RVE scale [2].

When using the computational homogenisation approach, if the behaviour of the RVE is not known in advance, the FE$^2$ homogenisation technique should be used, which involves attaching an RVE to each macroscopic integration point and solving a nonlinear BVP for each macroscopic integration point for each iteration of each step of the macroscopic problem. As a result, this would be computationally expensive. However, if the behaviour of the RVE can be predicted, the computational cost can be significantly reduced since there would be no need to solve the RVE BVP at each macroscopic integration point. As mentioned earlier, spiral strands exhibit nonlinear dissipative bending behavior when subjected to tensile force. Figure 1, shows the nonlinear bending response of a two-layer spiral strand with the geometric properties of Table 1 when subjected to an axial stress of 200 MPa. Due to the observed nonlinearity, predicting the bending behavior of RVE is not trivial. Therefore, the main objective of this study is to predict the cyclic response of a strand, using only its monotonic response through two different approaches, namely a rheological model and a data-based greybox model.

The paper is structured as follows. In Section 2, we present the rheological model employed to describe the inelastic bending response. Section 3 is dedicated to a brief introduction to a particular class of hysteresis operators, namely the Prandtl-Ishlinskii (P-I) operator formulated in terms of stops. After that, we show in Section 4 the equivalence of the two used methods, linking the weights and the thresholds characterising the P-I operators to the physical parameters of the spring system, namely spring stiffnesses and force thresholds. This reasoning allows us to relate the bending behaviour of cables to the parameters which uniquely identify a P-I operator and to find a physical interpretation for such quantities.

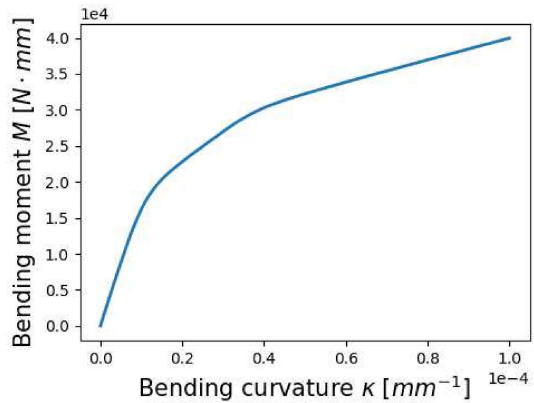

**Figure 1**. The monotonic bending response of the two-layer spiral strand under constant tensile force.

|        | No. of wires | Radius [mm] | Pitch length [mm] |
|--------|--------------|-------------|-------------------|
| Core   | 1            | 2.675       | -                 |
| Layer 1| 6            | 2.590       | 228.44            |
| Layer 2| 12           | 2.590       | 456.55            |

**Table 1**. Geometric properties of the two-layer spiral strand

## 2 RHEOLOGICAL MODEL

A typical cyclic bending response of a two-layer spiral strand subjected to constant tensile force is shown in Figure 2. To explain this behaviour, a typical cross section of a two-layer strand (Figure 2) is considered, and the contribution of the highlighted wire to the bending stiffness of the strand is examined. To this end, two different cases are considered. First, a case is considered where no tensile force is applied to the strand. In this case, the interlayer shear forces developed during the application of the bending curvature cause interlayer sliding, and therefore the wire bends about its own neutral axis, $x'$, instead of the cross-section neutral axis, $x$. Therefore, the only contribution from the wire, in this case, is $EI_{x'}$, where $E$ is the Young's modulus, and $I_{x'}$ is the second moment of area of the wire cross-section about the $x'$-axis. However, when tension is applied to the strand, interlayer frictional forces are developed. In this case, the frictional forces transfer the shear forces caused by bending between the wire and the first layer, and thus the wire bends about the $x$-axis. In this case, according to the parallel axis theorem, a secondary contribution, $\alpha EAd^2$, where $A$ is the cross-section of the wire and $d$ is the perpendicular distance between the $x$- and $x'$- axes, is also developed. However, when the shear forces exceed the frictional forces, the secondary contribution is lost. The multiplier $\alpha$ has been considered to account for the lay angle and the possibility of incomplete load transfer, especially for the wires with point interlayer contacts.

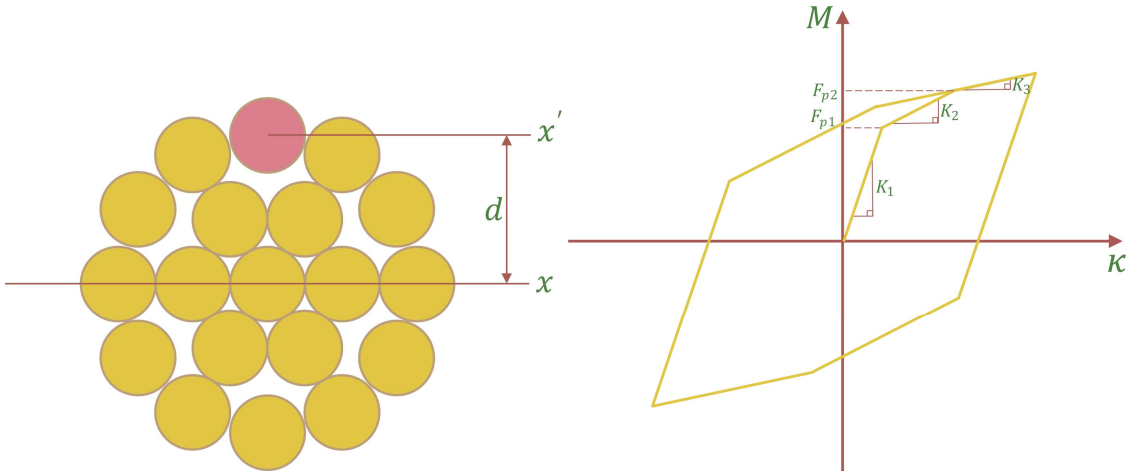

**Figure 2**. *Left*: A typical cross section of a two-layer spiral strand. *Right*: Schematic representation of the behaviour of a typical two-layer spiral strand under combined cyclic bending loading and constant tensile force.

Based on the above and by considering the schematic cyclic response of a two-layer strand (Figure 1), at low curvatures, the interlayer frictional forces are high enough to transfer the bending induced interlayer shear forces, and therefore, each layer has the primary and secondary contributions to the bending stiffness of the strand. Since these contributions are additive, they can be modeled using the so-called bending springs, where the force displacement is analogous to the moment curvature, acting in parallel. In this model, one spring represents the primary contribution from all wires and two springs represent the secondary contributions from the first and second layers. As the curvature increases, the secondary contribution from the second layer, followed by the first layer, is lost because the bending induced shear forces exceed the interlayer friction. Therefore, two sliders acting in series with the two springs representing the secondary contributions are used to model this loss. The rheological model representing a two-layer spiral strand is shown in Figure 3.

| $K_e^p$ [MPa] | $K_{p,1}^p$ [MPa] | $K_{p,2}^p$ [MPa] | $F_{p,1}^p$ [N.mm] | $F_{p,2}^p$ [N.mm] |
|---|---|---|---|---|
| 151942715 | 260046095 | 1340543977 | 10134 | 14338 |

**Table 2**. The parameters for the rheological model of the two-layer strand.

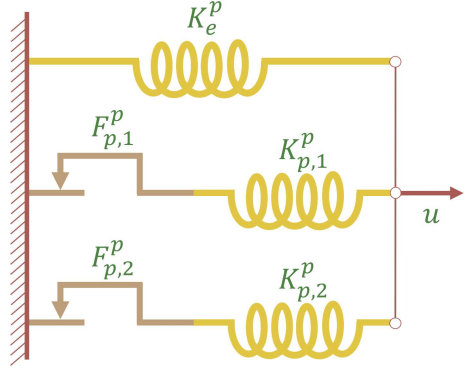

**Figure 3**. The rheological model representing the uniaxial bending response of two-layer strands.

According to Figure 1, the stiffness of the springs as well as the threshold of the sliders are easily obtained, from a monotonic bending response, as:

$$\begin{aligned}
K_e^p &= K_3, \\
K_{p,1}^p &= K_2 - K_3, \\
K_{p,2}^p &= K_1 - K_2, \\
F_{p,2}^p &= F_{p1} \frac{K_{p,2}^p}{K_1}, \\
F_{p,1}^p &= \left(F_{p2} - F_{p,2}^p\right) \frac{K_{p,1}^p}{K_2}.
\end{aligned} \quad (1)$$

Having the necessary parameters for the rheological models, the moment curvature response of the strand can be easily obtained by solving the equilibrium equations of the rheological model.

To illustrate the robustness of the proposed approach, the required parameters for the rheological model are extracted from the uniaxial bending response of the two-layer strand of Figure 1, and are given in Table 2. A comparison of the cyclic bending response obtained from the rheological model with the one obtained from homogenisation is given in Figure 4. It can be observed that although only a monotonic response has been used to identify the springs and sliders parameters, the cyclic response has been captured with very good accuracy using the proposed rheological model.

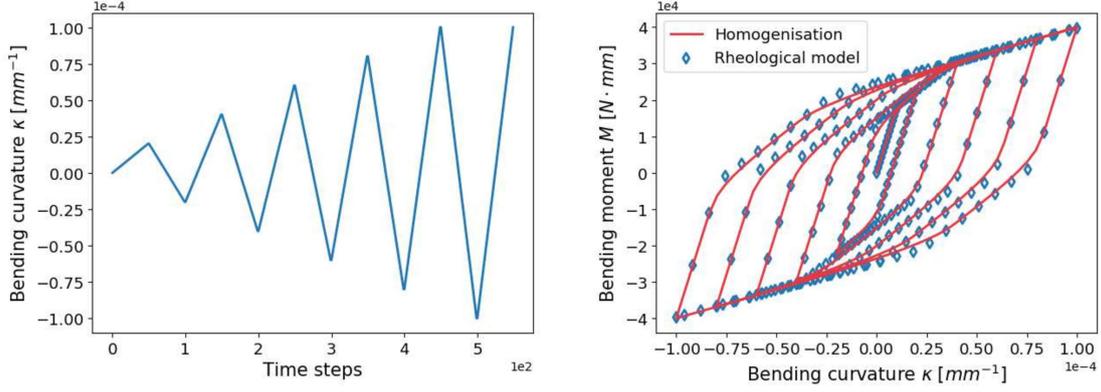

**Figure 4**. *Left*: The loading history. *Right*: A comparison of the cyclic uniaxial bending response of a two-layer spiral strand obtained from homogenisation and the rheological models.

## 3 PRANDTL-ISHLINSKII OPERATOR

As shown in [3, 4, 5], hysteresis operators are a well-studied topic with a variety of applications. The (P-I) operator $\mathscr{P}$ plays a relevant role in modelling the input-output relation in phenomena showing hysteretic behaviour and can be expressed as a superposition of elementary stop operators $\mathscr{S}_r$ multiplied by a suitable weight function $\omega(r)$, which is assumed to vanish for large values of $r$. We aim at expressing the bending moment ($M$) vs. bending curvature ($\kappa$) in terms of P-I operator as a discretised version of

$$M(t) = \mathscr{P}_r[\kappa](t) = \int_0^{+\infty} \omega(r)\mathscr{S}_r[\kappa](t)\mathrm{d}r. \qquad (2)$$

The P-I operator, as well as most hysteresis operators, are assumed to be rate-independent, i.e. the output is invariant with respect to changes of the time scale [5]. This means that we use the $t$ more as an "ordering parameter", rather than a time variable describing a dynamical process.

### 3.1 Stop operator

The stop operator $\mathscr{S}_r$ can be defined recursively, and can itself be equivalently formulated as a simple rheological model [3]. In particular, let $v$ be a generic piecewise monotone input, as the one shown in Figure 5 *left*. As long as the modulus of $v$ is smaller than the threshold $r$, the output $w$ is related to $v$ through a linear law with slope normalised to unity. Once the input has reached the yield value, the output remains constant under a further increase of the input; however, the linear behaviour is instantly recovered when the input is lowered again. As already mentioned, we write $w = \mathscr{S}_r[v]$.

The described behaviour of $\mathscr{S}_r$ can be described in analytic form for piecewise monotone inputs $v$ [5]. Suppose that $0 = t_0 < t_1 < ... < t_N = t_{\text{end}}$ is a partition of $[0, t_{\text{end}}]$ such that $v$ is monotone on each of the subintervals $[t_i, t_{i+1}]$ for $i = 0, ..., N-1$. Then $w = \mathscr{S}_r[v]$ is given by the inductive definition

$$\begin{aligned} & w(0) = s_r(v(0)), \\ & w(t) = s_r(v(t) - v(t_i) + w(t_i)) \quad \text{for } t_i < t \le t_{i+1}, \quad i = 0, ..., N+1 \\ & \text{with } s_r(v) = \min\{r, \max\{-r, v\}\}. \end{aligned} \qquad (3)$$

By superimposing different elementary stop operators, one is able to model more complex hysteretic effects, taking into account the history of the process. Figure 5 shows an example of input-output relation of the stop operator with threshold $r = 2$ applied to a sinusoidal signal. We also note that the weight function $\omega$ in (2) will characterise the linear behaviour of each contribution, by changing the value of the slope of the linear part of each single stop operator. From this, we already foresee that the weight function (or the discrete weights) will be closely related to the slopes of the diagram describing the constitutive response.

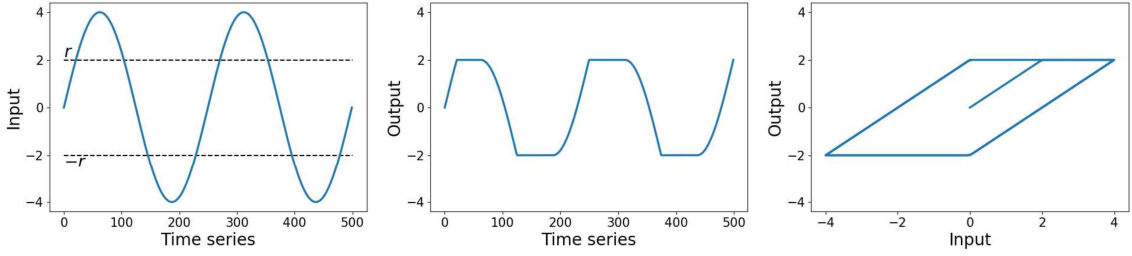

**Figure 5**. Example of stop operator with threshold $r = 2$. *Left*: Sinusoidal input function. *Centre*: Output function of the stop operator. *Right*: Input-output diagram.

## 4 P-I OPERATOR AND RHEOLOGICAL MODEL

We now want to find a suitable P-I operator able to describe the constitutive relation ($M$ vs. $\kappa$) that arises from the use of a spring model. By doing so, we will show the equivalence of the two methods, as well as find a physical interpretation to the weights and the threshold values of the chosen hysteresis operator.

From the shape of the diagram shown in Figure 1, we know that a discrete P-I operator defined as a superposition of 3 stop operators will be a suitable choice. We will then look for thresholds $r_i$ and weights $\omega_i$ for $i = 1, 2, 3$ such that

$$M(t) = \sum_{i=1}^{3} \omega_i \mathscr{S}_{r_i}[\kappa](t). \qquad (4)$$

Following the reasoning explained in section 3 and looking at definition (2), we deduce that

$$\begin{aligned} K_1 &= \omega_1 + \omega_2 + \omega_3 \\ K_2 &= \omega_2 + \omega_3 \\ K_3 &= \omega_3. \end{aligned} \qquad (5)$$

As anticipated, comparing (1) and (5), we see that the discrete weights can be found from the slope of the hysteresis diagram, and vice-versa, as

$$\omega_1 = K_{p_2}^p, \quad \omega_2 = K_{p_1}^p, \quad \omega_3 = K_e^p. \qquad (6)$$

In order to find the thresholds $r_i$ for $i = 1, 2, 3$, we think as follows. If we look at the initial branch of the diagram in Figure 1, we can say that the slope $m(\kappa) = K_1$ if $0 \leq \kappa < r_1$, $m(\kappa) = K_2$ if $r_1 \leq \kappa < r_2$ and $m(\kappa) = K_3$ if $r_2 \leq \kappa \leq r_3$. Taking this and (6) into account, by looking at the example diagram in Figure 5 *left*, we deduce that

$$r_i = \frac{F_{p,i}^p}{K_{p,i}^p} \quad \text{for } i = 1,2 \quad r_3 = \max_t(\kappa(t)). \qquad (7)$$

As shown in Figure 6, the stop operators $\mathscr{S}_{r_i}$ uniquely capture the contribution of different layers. In particular, we consider now a process in which the curvature increases monotonically from 0 to $r_3 = 10^{-4}$. For curvatures smaller than $r_1$, all contributions are active, hence the slope of the curve describing the strand response will be equal to the sum of all the positive slopes. This is equivalent to say that none of the stop operator has reached the yield value. For curvatures between the values of $r_1$ and $r_2$, the secondary contribution is lost, hence the slope of the red line equals the sum of the slopes of the black and green line. Equivalently, we can say that $\mathscr{S}_{r_1}[\kappa]$ remains constant when the input becomes greater than $r_1$. Finally, for curvatures bigger than $r_2$, also the secondary contribution of the first layer is lost, hence the strand behaves elastically and the only contribution left is the primary contribution of all wires.

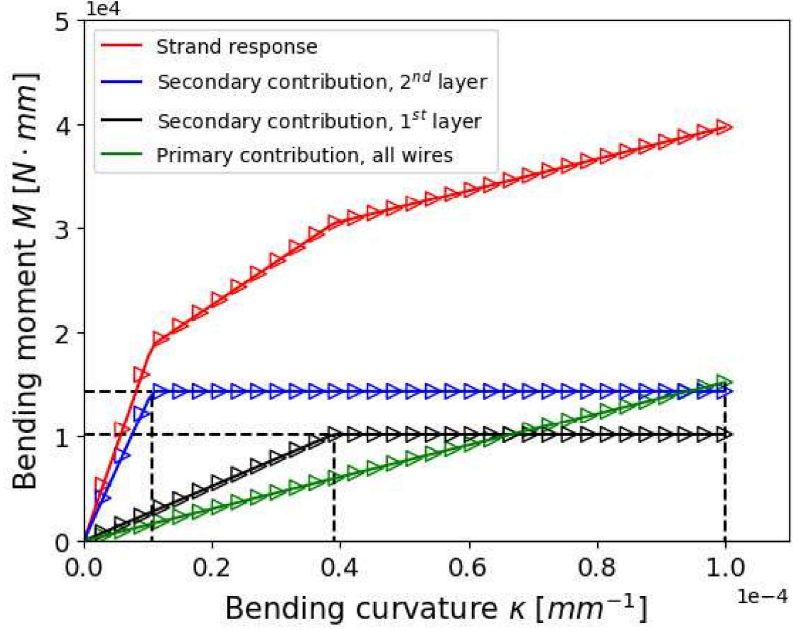

**Figure 6**. Different contributions of different layers. The solid lines represent the bending moment vs. bending curvature described by the rheological model, while the scattered plot are the branches of different stop operators multiplied by the corresponding weights. The perfect overlap of the plots is a further proof of the equivalence of the two models. One can note how the red line corresponds to the monotonic response shown in Figure 1. The horizontal dashed lines represent the force thresholds $F_{p,1}^p = 10134$ and $F_{p,2}^p = 14338$, while the vertical dotted lines corresponds to the thresholds related to the P-I operator, $r_1 = 1.069568 \times 10^{-5}$, $r_2 = 3.897001 \times 10^{-5}$, $r_3 = 10^{-4}$, see Table 2.

Figure 7 *right* compares the data obtained via the homogenisation technique and the P-I discrete operator with weights and thresholds defined as (6) and (7). As we can observe also in Figure 4 *right*, one is able to approximate very well the virtual measurements relative to the homogenisation approach with a rheological model composed by 3 springs, but of course one does not obtain an identical match. Analogously, this happens also if one uses a P-I operator formulated as a superposition of stops, as shown in Figure 7 *right*, where one can see a very good, but not identical, match between the data.

### 4.1 Weight function estimation

Moreover, given the thresholds defined as in (7) and the bending moment vs. bending curvature data coming from the virtual experiment based on the homogenisation procedure, we can approximate the discrete weights that will characterise a suitable P-I operator able to replicate the constitutive relation shown, for example, in Figure 2 *right*.

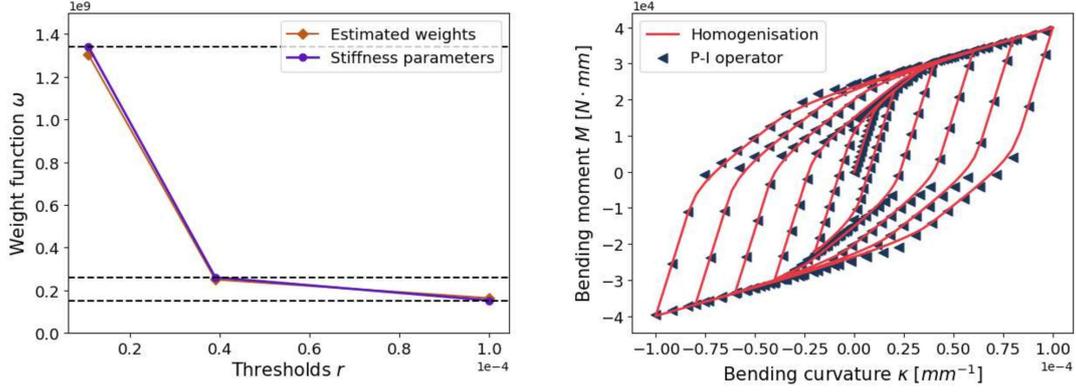

**Figure 7**. *Left*: Comparison between discrete approximated weights and discrete weights defined as spring stiffnesses. The dashed black horizontal lines represent, from top to bottom, $K_{p_2}^p$, $K_{p_1}^p$ and $K_e^p$, see Table 2. *Right*: Bending moment vs. bending curvature diagram: virtual experimental data and P-I operator.

Many techniques have been developed for such approximation, but in our case we proceed in a similar manner as in chapter 5 of [3]. More specifically, the approximated weights will result from the minimisation of a quadratic expression, constrained by a positivity condition. As one can see in 7 *left*, the estimated weights are in good agreement with the stiffness parameters values, and the difference between the values is due to the fact that the data coming from the homogenisation technique present a smooth transition between one regime to the other.

## 5 CONCLUSION

In this work, we have described two different techniques to describe the hysteretic inelastic behaviour of spiral strands subjected to bending deformation. First, a rheological model using the so-called bending springs and sliders has been used to describe the bending behaviour of spiral strands. The direct physical relevance of each element and the simple parameter identification process are the advantages of the proposed rheological model. Next, we make use of hysteresis operators of Prandtl-Ishlinskii type to describe such response. In particular, by superimposing simpler elements like the stop operators, we are able to build a model which is physically consistent and captures a complex inelastic behaviour. Even though this procedure might seem less intuitive than others, it allows us to describe the constitutive behaviour in a very compact manner. We have shown the equivalence of the two models (spring system and P-I operator) and the close relationship between their parameters.

Both methods can be used to model the bending behaviour of strands with an arbitrary number of layers, by adding more springs and sliders to the system shown in Figure 3 or superimposing a bigger number of stop operators to (4).

Future developments might foresee the extension of this comparison to biaxial bending, but most probably another class of hysteresis operators might be needed to model more complex phenomena.

### ACKNOWLEDGMENTS

This project has received funding from the European Union's Horizon 2020 research and innovation programme under the Marie Skłodowska-Curie grant agreement No 860124. 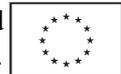

The authors would like to thank Klaus Kuhnen and Pavel Krejčí for fruitful in depth discussions on the mathematical theory and implementation of hysteresis operators.